\documentclass[floatfix,twocolumn]{aastex631}
\usepackage{threeparttable, longtable, threeparttablex, booktabs, url, amsmath}
\usepackage{verbatim}

\graphicspath{{./}{figures/}}
\RequirePackage[normalem]{ulem} 
\RequirePackage{color}\definecolor{RED}{rgb}{1,0,0}\definecolor{BLUE}{rgb}{0,0,1} 
\providecommand{\DIFaddbegin}{} 
\providecommand{\DIFaddend}{} 
\providecommand{\DIFdelbegin}{} 
\providecommand{\DIFdelend}{} 

\newcommand{\be}{\begin{eqnarray} }
\newcommand{\ee}{ \end{eqnarray} }

\shorttitle{An extremely active repeating fast radio burst source in a likely non-magneto-ionic environment}
\shortauthors{Feng et al.}
\graphicspath{{./}{figures/}}
\begin{document}

\title{An extremely active repeating fast radio burst source in a likely non-magneto-ionic environment}

\correspondingauthor{Yi Feng, D. Li, Y.-K. Zhang, Chao-Wei Tsai}
\email{yifeng@zhejianglab.com, dili@nao.cas.cn, ykzhang@nao.cas.cn, cwtsai@nao.cas.cn}

\author{Yi Feng}
\affil{Research Center for Astronomical Computing, Zhejiang Laboratory, Hangzhou 311100, China}
\affil{Institute for Astronomy, School of Physics, Zhejiang University, Hangzhou 310027, China}

\author[0000-0003-3010-7661]{Di Li}
\affil{National Astronomical Observatories, Chinese Academy of Sciences, Beijing 100101, China}
\affil{Research Center for Astronomical Computing, Zhejiang Laboratory, Hangzhou 311100, China}
\affil{University of Chinese Academy of Sciences, Beijing 100049, China}

\author{Yong-Kun Zhang}
\affil{National Astronomical Observatories, Chinese Academy of Sciences, Beijing 100101, China}
\affil{University of Chinese Academy of Sciences, Beijing 100049, China}

\author[0000-0002-9390-9672]{Chao-Wei Tsai}
\affil{National Astronomical Observatories, Chinese Academy of Sciences, Beijing 100101, China}
\affiliation{Institute for Frontiers in Astronomy and Astrophysics, Beijing Normal University,  Beijing 102206, China}
\affiliation{University of Chinese Academy of Sciences, Beijing 100049, China}

\author[0000-0003-4721-4869]{Yuanhong Qu}
\affil{Nevada Center for Astrophysics, University of Nevada, Las Vegas, NV 89154, USA}
\affil{Department of Physics and Astronomy, University of Nevada, Las Vegas, NV 89154, USA}

\author{Wei-Yang Wang}
\affil{University of Chinese Academy of Sciences, Beijing 100049, China}
\affil{Department of Astronomy, Peking University, Beijing 100871, China}
\affil{Kavli Institute for Astronomy and Astrophysics, Peking University, Beijing 100871, China}

\author{Yuan-Pei Yang}
\affil{South-Western Institute for Astronomy Research, Yunnan University, Kunming 650504, China}
\affil{Purple Mountain Observatory, Chinese Academy of Sciences, Nanjing 210023, China}

\author{Pei Wang}
\affil{National Astronomical Observatories, Chinese Academy of Sciences, Beijing 100101, China}
\affil{Institute for Frontiers in Astronomy and Astrophysics, Beijing Normal University,  Beijing 102206, China}

\author{Dengke Zhou}
\affil{Research Center for Astronomical Computing, Zhejiang Laboratory, Hangzhou 311100, China}

\author{Jiarui Niu}
\affil{National Astronomical Observatories, Chinese Academy of Sciences, Beijing 100101, China}

\author{Chenchen Miao}
\affil{National Astronomical Observatories, Chinese Academy of Sciences, Beijing 100101, China}

\author{Mao Yuan}
\affil{National Astronomical Observatories, Chinese Academy of Sciences, Beijing 100101, China}

\author{Jiaying Xu}
\affil{Research Center for Astronomical Computing, Zhejiang Laboratory, Hangzhou 311100, China}

\author{Ryan S. Lynch}
\affil{Green Bank Observatory, Green Bank, WV, 24944, USA}

\author[0000-0002-7045-9277]{Will Armentrout}
\affil{Green Bank Observatory, Green Bank, WV, 24944, USA}

\author{Brenne Gregory}
\affil{Green Bank Observatory, Green Bank, WV, 24944, USA}

\author{Lingqi Meng}
\affil{National Astronomical Observatories, Chinese Academy of Sciences, Beijing 100101, China}

\author{Shen Wang}
\affil{School of Computer Science, Fudan University, Shanghai 200433, China}

\author{Xianglei Chen}
\affil{National Astronomical Observatories, Chinese Academy of Sciences, Beijing 100101, China}

\author{Shi Dai}
\affil{School of Science, Western Sydney University, Locked Bag 1797, Penrith NSW 2751, Australia}

\author{Chen-Hui Niu}
\affil{Institute of Astrophysics, Central China Normal University, Wuhan 430079, Hubei, China}
\affil{National Astronomical Observatories, Chinese Academy of Sciences, Beijing 100101, China}

\author{Mengyao Xue}
\affil{National Astronomical Observatories, Chinese Academy of Sciences, Beijing 100101, China}

\author{Ju-Mei Yao}
\affil{Xinjiang Astronomical Observatory, Chinese Academy of Sciences, Urumqi, Xinjiang 830011, China}

\author{Bing Zhang}
\affil{Nevada Center for Astrophysics, University of Nevada, Las Vegas, NV 89154, USA}
\affil{Department of Physics and Astronomy, University of Nevada, Las Vegas, NV 89154, USA}

\author{Junshuo Zhang}
\affil{National Astronomical Observatories, Chinese Academy of Sciences, Beijing 100101, China}

\author{Weiwei Zhu}
\affil{National Astronomical Observatories, Chinese Academy of Sciences, Beijing 100101, China}
\affil{Institute for Frontiers in Astronomy and Astrophysics, Beijing Normal University,  Beijing 102206, China}

\author{Jintao Xie}
\affil{Research Center for Astronomical Computing, Zhejiang Laboratory, Hangzhou 311100, China}

\author{Yuhao Zhu}
\affil{National Astronomical Observatories, Chinese Academy of Sciences, Beijing 100101, China}
\affil{University of Chinese Academy of Sciences, Beijing 100049, People's Republic of China}



\begin{abstract}
Fast radio bursts (FRBs) are bright radio bursts originating at cosmological distances. Only three repeating FRBs FRB~20121102A, FRB~20190520B and FRB~20201124A among $\sim 60$ known repeating FRBs have circular polarization.
 We observed the FRB 20220912A with the Robert C. Byrd Green Bank Telescope (GBT) at L-band on 24 October 2022 and detected 128 bursts in 1.4 hours, corresponding to a burst rate of about 90 hr$^{-1}$, which is the highest yet for FRBs observed by the GBT. The average rotation measure (RM) was $-$0.4$\pm$0.3\,rad\,m$^{-2}$ with negligible intraday RM change, indicating a likely non-magneto-ionic environment.  
61\% bursts have linear polarization fraction greater than 90\%. Approximately 56\% of the bright bursts have circular polarization. 
A downward drift in frequency and polarization angle 
swings were  found in our sample. The characterization of FRB~20220912A indicates that the circular polarization is unlikely to be caused by the magneto-ionic environment for at least some of the repeating FRB population.

\end{abstract}



\section{Introduction} \label{sec:intro}
Fast radio bursts (FRBs) are bright radio transients first discovered by \cite{2007Sci...318..777L}. While their cosmological origin and energetic nature make them ideal tools for probing cosmic web, Galactic haloes, baryons, etc. (\citealt{2016Sci...354.1249R}, \citealt{2019Sci...366..231P}, \citealt{2020Natur.581..391M}), their progenitors and radiation mechanisms are still unknown. A particularly interesting subset of FRBs is the so-called repeating FRBs, which recurrently emit radio bursts. Among $\sim 800$ FRBs, $\sim 60$ are repeating FRBs\footnote{https://blinkverse.alkaidos.cn/} \citep{blinkverse}. 

High levels of activity from a repeating FRB source discovered
by Canadian Hydrogen Intensity Mapping Experiment (CHIME), FRB~20220912A, was announced in
October 2022 \citep{2022ATel15679....1M}. \cite{2022ATel15679....1M} reported nine bursts from FRB~20220912A in three days of observations and estimated a burst rate as high as 200-300 bursts
per day at $\sim$\,Jy ms fluence thresholds. FRB~20220912A has a dispersion measure (DM) of 219.46 \,pc~cm$^{-3}$ \citep{2022ATel15679....1M}. The maximum Galactic DM contribution along this line of sight is about 125 and 122 \,pc~cm$^{-3}$ for NE2001 model \citep{2002astro.ph..7156C} and YMW16 model \citep{2017ApJ...835...29Y} , respectively. This active repeating FRB was then followed up by telescopes world-wide, including the Deep Synoptic Array (DSA-110) \citep{2022ATel15716....1R}, the Big Scanning Antenna (BSA) \citep{2022ATel15713....1F}, the Stockert telescope \citep{2022ATel15691....1H}, the Robert C. Byrd Green Bank Telescope (GBT) \citep{2022ATel15723....1F}, the Effelsberg \citep{2022ATel15727....1K}, the Five-hundred-meter Aperture Spherical radio Telescope (FAST) \citep{2022ATel15733....1Z}, the Arecibo 12-m \citep{2022ATel15734....1P}, the Allen Telescope Array \citep{2022ATel15735....1S}, the Tianlai \citep{2022ATel15758....1Y}, the Nancay radio telescope (NRT) \citep{2023MNRAS.526.2039H}, and the upgraded Giant Metrewave Radio Telescope (GMRT) \citep{2022ATel15806....1B}. DSA-110 localized FRB~20220912A to 
Right Ascension (RA) 23$^h$09$^m$04.9$^s$, Declination (Dec) +48$^\circ$42$'$25.4$''$ (J2000 equinox), consistent with the galaxy PSO~J347.2702+48.7066 \citep{0912red} at a redshift of 0.0771.

High levels of activity has also been observed in other repeating FRB sources. For example, FRB~20121102A and FRB~20201124A had peak burst rate of 122 hr$^{-1}$ \citep{li21} and  542 hr$^{-1}$ \citep{2022RAA....22l4002Z}, respectively. Both rate measurements were obtained by the FAST (\citealt{2011IJMPD..20..989N}, \citealt{2018IMMag..19..112L}). The burst rate of the overall repeater population typically ranges from 0.1 hr$^{-1}$ to $\sim$10 hr$^{-1}$ (\citealt{chime8}, \citealt{chime9}, \citealt{luo2020}, \citealt{chime25}, \citealt{kumar2023}). We caution that the burst rate of some repeaters could be higher because a fraction of weak bursts cannot be detected. FRB 20121102A and FRB 20201124A are likely located in complex environments. FRB~20121102A has a large rotation measure (RM) scatter ($\sigma_{\mathrm{RM}}$) of $30.9\,\mathrm{rad/m}^2$ \citep{feng22}, which has been attributed to multi-path scattering arising from complex environment. The RM of FRB~20121102A decreased from $1.03\times10^5$\,rad\,m$^{-2}$ to $6.7\times10^4$\,rad\,m$^{-2}$ between January 2017 and August 2019 (\citealt{121102rm}, \citealt{2021ApJ...908L..10H}). The RM decreased to $3.1\times10^4$\,rad\,m$^{-2}$  in 2023 \citep{2023ATel15980....1F}, continuing the decreasing trend. 
The large RM scatter and variable RMs of FRB~20121102A indicate that there are some complex dynamic magneto-ionic environments near the FRB source, which may be related to supernova remnants (SNR) or wind nebulae (\citealt{2018ApJ...861..150P}, \citealt{2018ApJ...868L...4M}, \citealt{2021ApJ...923L..17Z}, \citealt{2022MNRAS.510L..42K},
\citealt{2022ApJ...928L..16Y},
\citealt{2023MNRAS.520.2039Y}). 
FRB~20201124A shows irregular short-time variation of the RM followed by a 
constant RM \citep{2022Natur.609..685X}, suggesting a complicated, dynamically evolving, magnetized immediate environment. A similar behavior (irregular RM variation followed by a constant RM) has been observed in pulsar binary systems, which implies some FRB source might be in a binary system with an elliptical orbit (\citealt{2022NatCo..13.4382W}, \citealt{2023ApJ...942..102Z}, \citealt{2023MNRAS.520.2039Y}).

Unlike FRB~20121102A and FRB~20201124A, here we report the likely non-magneto-ionic environment of the active repeating FRB~20220912A.  
Our observations and data processing procedures are described in Section~\ref{sec:methods}. Our results are presented in Section~\ref{sec:re}. 
We discuss the results and conclude in Section~\ref{sec:cir}.

\section{methods} \label{sec:methods}
\subsection{observations and burst search} \label{sec:data}
FRB~20220912A was observed between 1.1--1.9 GHz with the GBT's L-Band receiver and the Versatile Green Bank Astronomical Spectrometer (VEGAS) digital backend \citep{2015ursi.confE...4P} on 24 October 2022 starting at
00:30:46 UTC and lasting 2 hours. After setting up and calibration, the on-source time was 1.4 hours. The data were coherently dedispersed at a DM of 219.46 \,pc~cm$^{-3}$ and recorded in the \textsc{psrfits} standard format \citep{2004PASA...21..302H}. Full-Stokes spectra were recorded every 81.92 $\mu$s with 0.195 MHz-wide channels.

Pulse searching is grounded in our developed deep learning-based search tool (Zhang et al, in prep), named DRAFTS \footnote{\url{https://github.com/SukiYume/DRAFTS}}. Here, we employ a classification model based on ResNet50V2 \citep{he2016deep, he2016identity} from the DRAFTS framework, which has been well-trained on real FRB data detected by FAST. The model receives time-frequency data after dispersion correction, with an input size of $1\times512\times512$, and outputs the probability (ranging from 0 to 1) of an FRB being present in the data. For the GBT data, we apply dispersion compensation with  DM = $219.46\,\rm pc\,cm^{-3}$, followed by fourfold data subsampling. The data are then divided into distinct time segments, each consisting of 4096 samples (post-subsampling), and reshaped into a size of $512\times512$. These segments are subsequently fed into the model for prediction. When the probability of a signal in the model-predicted data segment exceeds 50\% (as the output of the softmax activation function $>0.5$), we retain this data segment as a signal. This criterion differs from the traditional signal-to-noise ratio definition. Ultimately, 128 bursts are preserved.


\subsection{flux and polarization calibration} \label{sec:cal}
Data were calibrated using the \textsc{psrchive} package \cite{2004PASA...21..302H}.  On and off-source observations of the standard calibrator 3C 380 were used to measure the intensity of the L-Band receiver's built-in noise diode.  The noise diode was observed again at the position of FRB~20220912A prior to the main observations, and these data were used to flux calibrate each burst from the FRB.

Polarization calibration was achieved by correcting for the differential gain and phase between the receptors through separate measurements of a noise diode signal injected at an angle of $45^{\circ}$ from the linear receptors with the single-axis model using the PSRCHIVE software package. 
To excise radio frequency interference (RFI), we used the PSRCHIVE software package to adopt median filter to each burst in the frequency domain with the command paz and we also mitigated RFI of each burst manually with the command pazi. 

We measured the RM using the methods of Stokes QU-fitting \citep{2012MNRAS.421.3300O}. We selected positions where the signal exceeds the noise by 3 sigma for RM fitting. For each burst, we fitted an average RM. We calculated the degrees of linear polarization and circular polarization for bursts with RM detection. We first derotated the linear polarization with the measured RM. The measured linear polarization is overestimated in the presence of noise. Therefore we use the frequency-averaged, de-biased total linear polarization \citep{2001ApJ...553..341E,askap20}\footnote{A typographical error in \cite{2001ApJ...553..341E} was corrected in \cite{askap20}.} :
\begin{equation} \label{eq:L_de-bias}
    L_{{\mathrm{de\mbox{-}bias}}} =
    \begin{cases}
      \sigma_I \sqrt{\left(\frac{L_{i}}{\sigma_I}\right)^2 - 1} & \text{if $\frac{L_{i}}{\sigma_I} > 1.57$} \\
      0 & \text{otherwise} ,
    \end{cases}
\end{equation}
where $\sigma_I$ is the Stokes I off-pulse standard deviation and $L_i$ is the measured frequency-averaged linear polarization of time sample $i$.
We defined the degree of linear polarization 
as ($\Sigma_{i} L_{\mathrm{de\mbox{-}bias},i}$)/($\Sigma_{i}I_i$)
and that of circular polarization as ($\Sigma_{i} V_i$)/($\Sigma_{i}I_i$), where the summation is over the time samples in one burst and $V_i$ is the measured frequency-averaged circular polarization of time sample $i$. 
Defining $I = \Sigma_{i}I_i$, $L = \Sigma_{i} L_{\mathrm{de\mbox{-}bias},i}$ and $V = \Sigma_{i}V_i$, uncertainties on the linear polarization fraction and circular polarization fraction are calculated as:
\begin{equation} \label{eq:uncertainty}
    \sigma_{\rho/I} = \frac{\sqrt{N+N\frac{\rho^2}{I^2}}}{I}\sigma_{I},
\end{equation}
where $N$ is the number of time samples of the burst (signal exceeds the noise by 3$\sigma$), and $\rho = L,V$ for linear and circular polarization fraction, respectively.

\subsection{$\textsc{DM}$, fluence and energy} \label{sec:energy}

We use \textsc{DM\_phase} \footnote{\url{https://github.com/DanieleMichilli/DM\_phase}} for DM optimization. After the flux calibration, we calculated the peak flux density $S_{\rm peak}$ and fluence $F$ from the data. Due to the presence of RFI in many bursts, we determined the upper and lower cutoff frequencies (bandwidth) where the signal is submerged in noise, through visual inspection. The choice of bandwidth does not affect the energy calculation as long as it adequately covers the burst. In the case of bursts with substructure, we defined the burst bandwidth based on the widest structure within it. The pulse profile was obtained by averaging the flux in burst bandwidth along the frequency. The peak flux is the maximum value of the pulse profile with the time resolution of 655.36 $\mu$s. Specific fluence is the integration of the burst profile with respect to time, and the equivalent width $W_{\rm eq}$ is computed by dividing the specific fluence by the peak flux.

We calculate the isotropic equivalent burst energy following the Equation:
\begin{equation}\label{eq:g}
    E = 10^{39} {\rm erg}\frac{4\pi}{1+z} \left(\frac{D_L}{10^{28}{\rm cm}}\right)^{2} \left(\frac{F}{\rm Jy\cdot ms}\right)\left(\frac{\Delta\nu}{\rm GHz}\right),
\end{equation}
where $F=S_{\rm peak}\times W_{\rm eq}$ is the specific fluence, $\Delta \nu$ is the burst bandwidth. The adoption of $\Delta \nu$ is more relevant for narrow-band FRBs, which is typically the case for the bursts in repeaters. $D_L=360.86\,\rm Mpc$ is the luminosity distance of FRB~20220912A corresponding to the redshift $z=0.0771$ \citep{0912red} adopting the standard Planck cosmological model \citep{2016A&A...594A..13P}.

\section{results} \label{sec:re}
We detected 128 bursts from FRB~20220912A in 1.4 hours with the GBT (see Methods~\ref{sec:data} for details of the observation). The time of arrivals, peak flux density, width, fluence, energy, DM, RM and degrees of linear and circular polarization of the bursts are listed in Table~\ref{tab:burst}. The average burst rate is about 90 hr$^{-1}$, which is the highest for fast radio burst observations at the GBT or any other (sub)-100 meter radio telescopes. The previous highest burst rate at the GBT is 21 hr$^{-1}$ of FRB~20121102A \citep{2018ApJ...863....2G}. \cite{nimmo23} reported a burst storm of 53 bursts occuring within only 40 minutes from FRB~20200120E at Effelsberg telescope, which corresponds to a burst rate of 80 hr$^{-1}$.   

\begin{figure}
\includegraphics[scale=0.65]{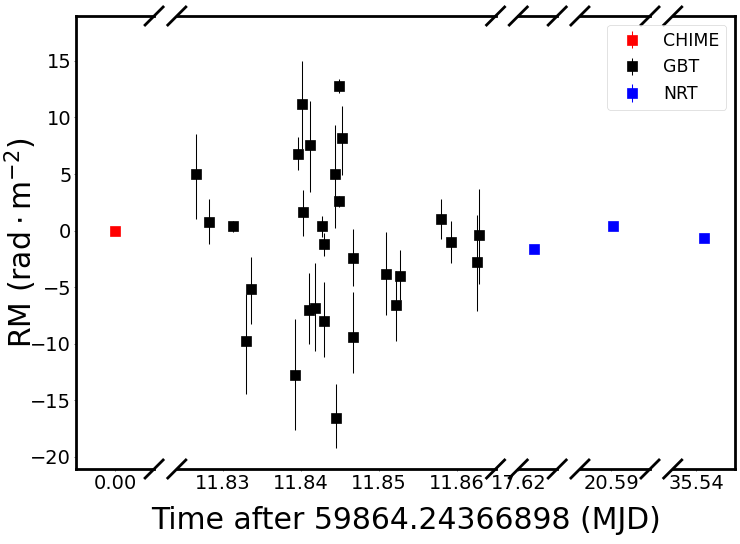}
\caption{RM evolution of FRB~20220912A. Different markers indicate at which telescope the burst was detected.}
\label{rm}
\end{figure}

Active repeating FRBs~20121102A and 20201124A show signs of being in complex environments. Although FRB~20220912A is extremely active, the environment of FRB~20220912A seem to be non-magneto-ionic. We show the RM of FRB~20220912A in Figure~\ref{rm}. The RMs have been corrected for the effect of the ionosphere. We calculate the ionosphere's RM by using a python code - \texttt{IonFR} \citep{sotomayor2013calibrating} and global ionospheric map (GIM) products. The GIM products are available on NASA's website\footnote{cddis.nasa.gov/archive/gnss/products/ionex/} in the IONosphere EXchange (IONEX) format.
For our GBT observation, the ionosphere contribution is $1.69\pm0.06$\,rad\,m$^{-2}$. In Figure \ref{rm}, we only selected bursts with RM error bars smaller than 5\,rad\,m$^{-2}$. We note that burst 89 has a RM of 14.5$\pm$0.6\,rad\,m$^{-2}$. This RM measurement is an outlier and is caused by frequency-dependant polarization. Excluding this measurement and bursts with RM error bars larger than 5\,rad\,m$^{-2}$, our GBT measurements have a mean RM of $-$0.4$\pm$0.3\,rad\,m$^{-2}$.
We did not observe obvious intraday RM change in FRB~20220912A in contrast to other active repeating FRBs. FRB~20121102A shows intraday RM change of a few hundred rad\,m$^{-2}$ \citep{2021ApJ...908L..10H}. Similar to FRB~20121102A, FRB~20190520B also shows intraday RM change of a few hundred rad\,m$^{-2}$ \citep{reshma22}. FRB~20201124A has a smaller intraday RM change of a few tens rad\,m$^{-2}$ \citep{2022Natur.609..685X}. In Figure~\ref{rm}, we also show the RM measurements from CHIME \citep{2022ATel15679....1M} and NRT \citep{2023MNRAS.526.2039H}. 
The RM does not change in about a month's time. The estimated Galactic RM contribution in this line of sight is $-$15$\pm$11\,rad\,m$^{-2}$ \citep{2022A&A...657A..43H}. The local environment of FRB~20220912A thus contributes little RM. The small and stable RMs indicate a non-magneto-ionic environment of FRB~20220912A. The observation of FRB~20201124A implies that the RM variation could be intermittent for a certain source, meanwhile, the small mean value might be due to the cancellation of RM at different region. Thus, in the future we will conduct a long-term monitoring campaign for FRB~20220912A, which would reveal whether such a source indeed has a non-magneto-ionic environment. 

Out of the 128 detected bursts, 61\% bursts have linear polarization fraction greater than 90\%. For bursts with peak flux density greater than 500\,mJy, 56\% of the bursts have circular polarization with signal-to-noise ratio $>$ 5.
The highest absolute fractional circular polarization reaches 58\%.  
FRB~20220912A is the 4th reported repeating FRB with circular polarization. The previous reported repeating FRBs with circular polarization are all active FRBs, namely FRB~20201124A \citep{2022Natur.609..685X}, FRB~20190520B and FRB~20121102A \citep{2022SciBu..67.2398F}. The fraction of bursts with circular polarization is less than 10\% for FRB~20190520B and FRB~20121102A \citep{2022SciBu..67.2398F}, much smaller than FRB~20220912A. 

\begin{figure*}[!htp]
  \centering
  \includegraphics[width=0.46\textwidth]{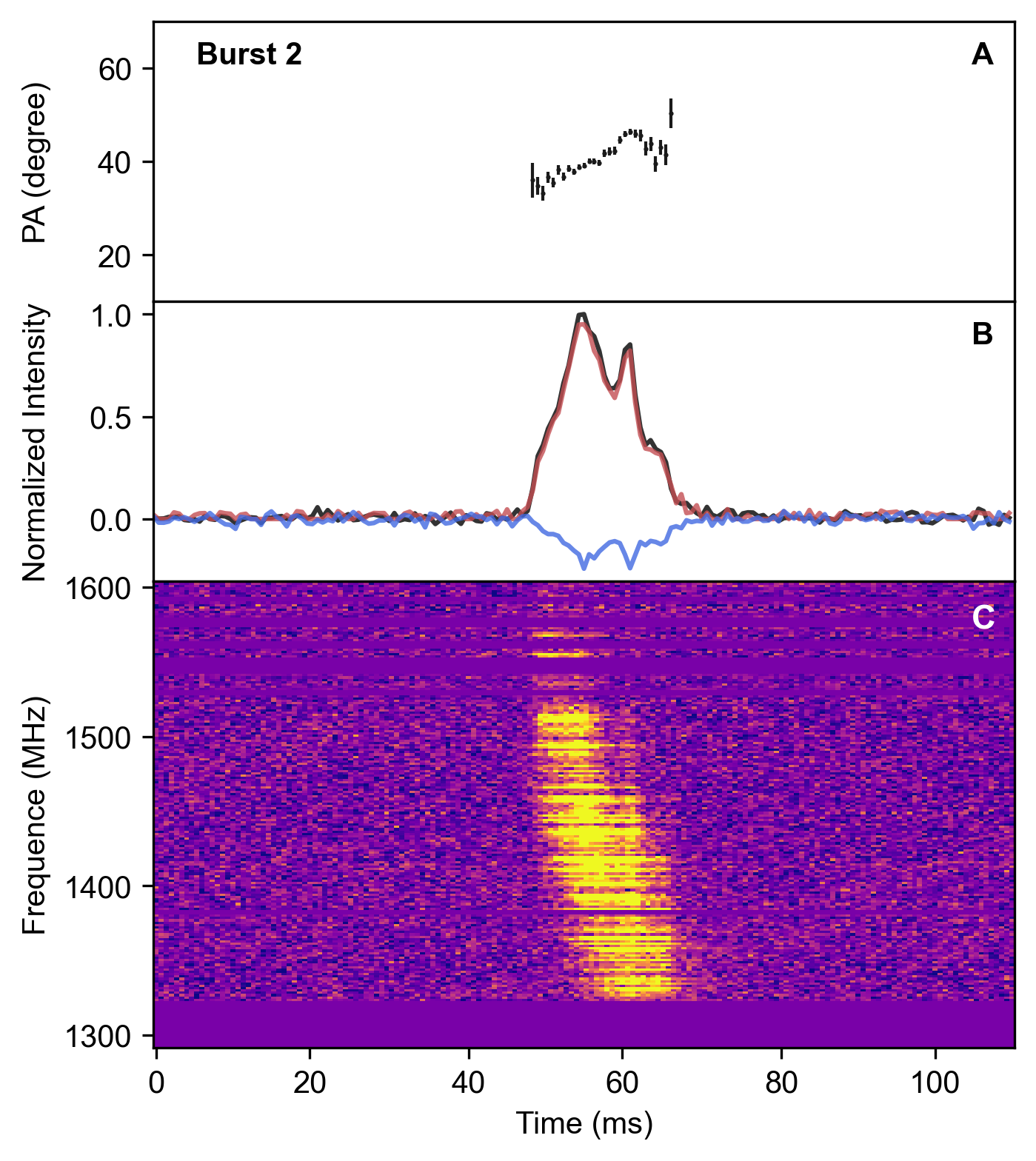}
  \includegraphics[width=0.46\textwidth]{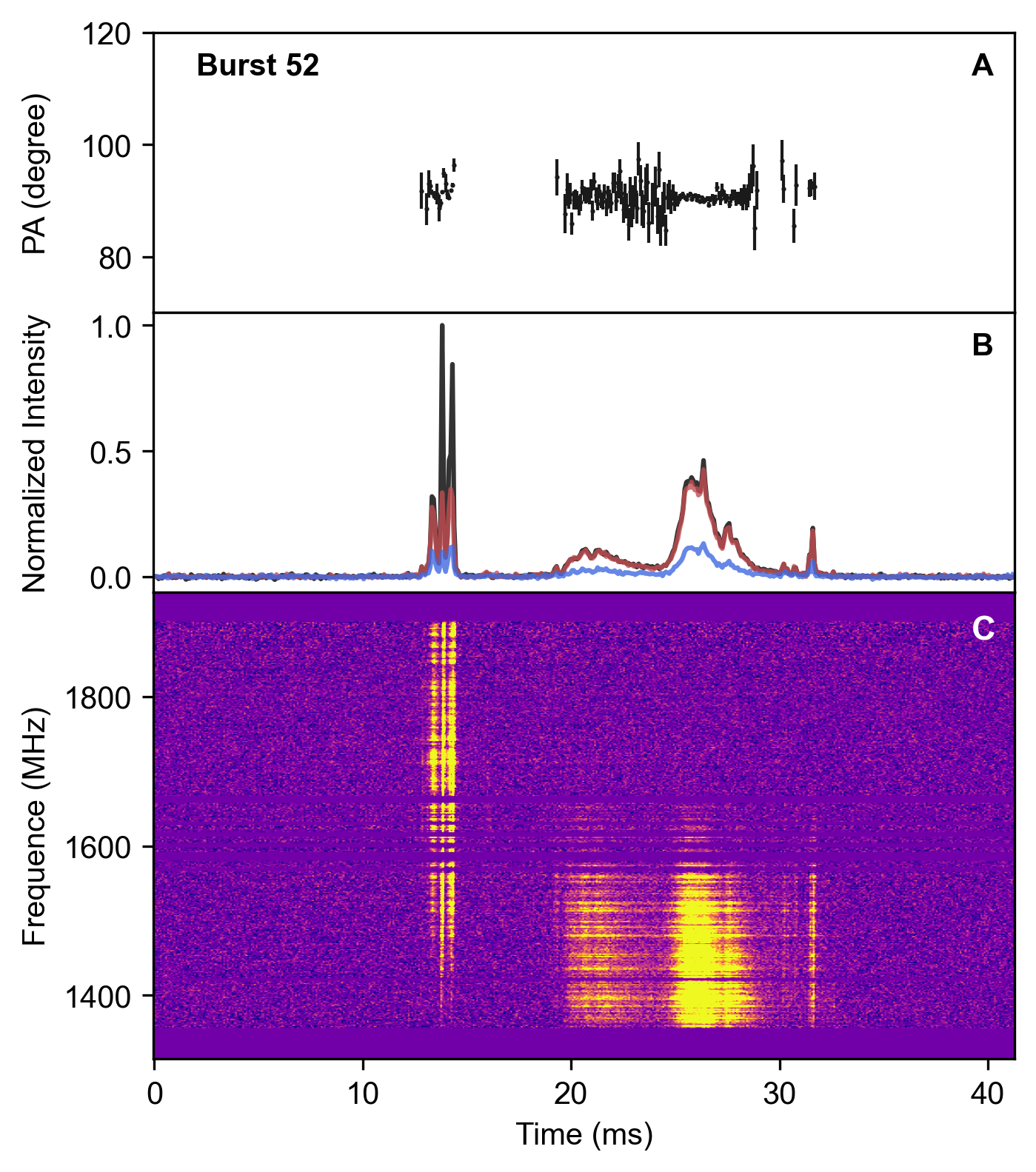}\\
  \includegraphics[width=0.46\textwidth]{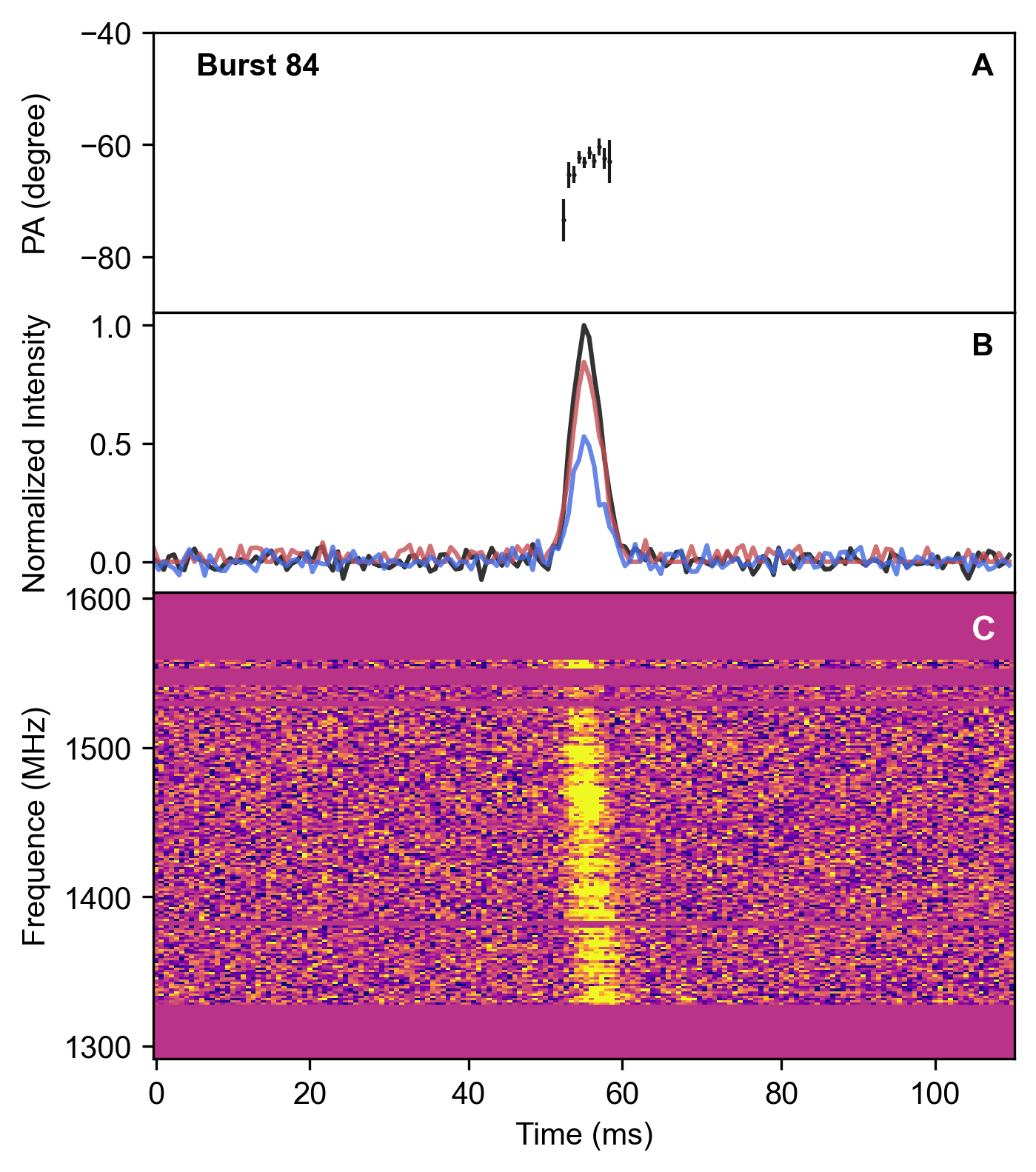}
  \includegraphics[width=0.46\textwidth]{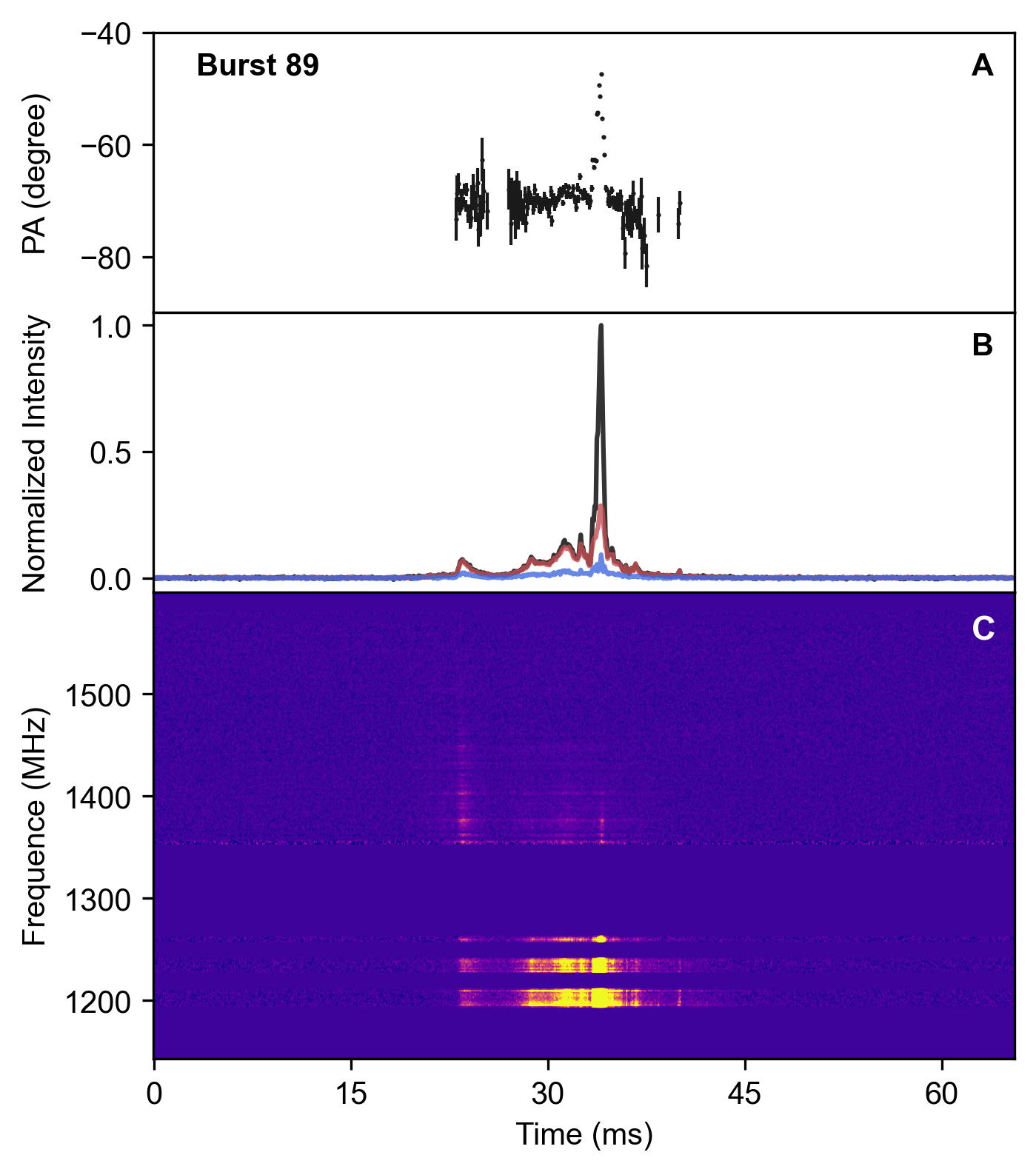}
  \caption{Polarization position angle and intensity profiles with dynamic spectra of four bursts of FRB~20220912A. In each panel, sub-panel A shows the polarization position angles; sub-panel B shows the polarization pulse profiles with lines indicating total intensity (black, normalized to a peak value of 1.0), linear polarization (red) and circular polarization (blue); sub-panel C shows the dynamic spectra.}\label{fig:spec}
\end{figure*}

Polarization angle swings are observed in some bursts of FRB~20220912A. The other repeating FRBs with polarization angle swings include FRB~20180301A \citep{luo2020} and FRB~20201124A \citep{2022Natur.609..685X}. In Figure~\ref{fig:spec}, we show polarization angles, pulse profiles and dynamic spectra of a few representative bursts.
As shown in the figure, bursts 2 and 89 have clear polarization angle swings. We also note that burst 52 shows downward drift in frequency \citep{2019ApJ...876L..23H}. 

\section{DISCUSSIONS AND CONCLUSIONS}\label{sec:cir}
Radiation mechanisms for repeating FRBs can be generally grouped into inside \citep{Katz2014,Kumar2017,Yang&Zhang2018,Lu2020,Cooper2021,Zhang2022,Qu&Zhang2024} and outside the magnetar magnetosphere \citep{Lyubarsky2014,Metzger2019,Plotnikov,Beloborodov2020,Sironi2021}.
Observationally, most bursts have nearly 100\% linear polarization and some bursts can have high circular polarization, which might be produced by intrinsic radiation mechanisms and propagation effects inside or outside the magnetosphere \citep{Qu&Zhang2023}. 
Inside the magnetosphere, such a highly linearly polarized wave can be produced via on-axis curvature radiation by charged bunches \citep{2022ApJ...927..105W,2022RAA....22g5013T}, and no obvious trend of sign change of circular polarization in bursts suggest that most emitting bunches have opening angels larger than $1/\gamma$ \citep{2022MNRAS.517.5080W}, where $\gamma$ is the electron Lorentz factor. 
Another possible radiation mechanism is the coherent inverse Compton scattering (ICS) by charged bunches \citep{Zhang2022,Qu&Zhang2024}. The ICS radiation is nearly 100\% linearly polarized for the on-axis case. Due to the different phases and different polarization angles of ICS radiation, circular polarization can be produced in an off-axis geometry when a geometric bunch is considered \citep{Qu&Zhang2023}. 
Generally speaking, for the two magnetospheric mechanisms, the radiation could be highly circular polarized even if the line of sight is within the radiation cone ($<1/\gamma$). 
Outside the magnetosphere, synchrotron maser mechanism in an ordered magnetic field has been discussed to power FRBs, the radiation is highly linearly polarized when X-mode dominates with magnetization $\sigma\gg1$ \citep{Plotnikov,Sironi2021}.

Most bursts have 100\% linear polarization which implies a complex configuration rather than a single charged bunch. 
The bunch cross section could be large enough across a bundle of open magnetic field line region. In such a case, high linear polarization is still sustained when the line of sight is within the opening angle of the bundle. 
The non-magneto-ionic environment suggests that some propagation effects are not dominant, e.g., the synchrotron absorption effect in the nebula cannot exist. Thus the high circular polarization might be mainly produced inside the magnetosphere. When the background pair plasma has asymmetric distribution of Lorentz factor, the circular polarization might be generated through cyclotron resonance absorption for a slowly rotating ($\geq3$ s) magnetar with lower surface magnetic field strength ($\leq10^{14}$ G) \citep{Qu&Zhang2023}. Otherwise, intrinsic radiation mechanisms are required to produce circular polarization for an off-axis case.

We report the extreme activity of FRB~20220912A and its likely non-magneto-ionic environment. Our main results are the following.
\begin{itemize}
\setlength{\itemsep}{3pt}
 \item[1.] 128 bursts were detected in 1.4 hours with GBT, corresponding to a burst rate of about 90 hr$^{-1}$, which is the highest for any FRB observed by GBT or any other (sub) 100-m radio telescopes.

\item[2.] It has an average RM of -0.4$\pm$0.3\,rad\,m$^{-2}$ with unnoticeable intraday RM change, which indicates that it is in a non-magneto-ionic environment.  

\item[3.] Out of the 128 detected bursts, 61\% bursts have linear polarization fraction greater than 90\%.  56\% of the bright bursts have circular polarization. 

\item[4.] Some bursts from FRB~20220912A show downward drift in frequency and polarization angle swings. Our results have increased the number of repeating FRBs with circular polarization from three to four. \end{itemize} 

The likely non-magneto-ionic environment of FRB~20220912A and characterization of FRB~20220912A indicates that the circular polarization is unlikely caused by the magneto-ionic environment, but more likely related to the central engine, for at least some of the repeating FRB population.

\begin{acknowledgments}
This work is supported by National Key R\&D Program of China No. 2023YFE0110500, National Natural Science Foundation of China grant No.\ 11988101, 12203045, 11725313, by the Leading Innovation and Entrepreneurship Team of Zhejiang Province of China grant No. 2023R01008, and by Key R\&D Program of Zhejiang grant No. 2024SSYS0012. Yuan-Pei Yang is supported by National Natural Science Foundation of China grant No. 12003028 and the China Manned Spaced Project (CMS-CSST-2021-B11). The Green Bank Observatory is a facility of the National Science Foundation operated under cooperative agreement by Associated Universities, Inc.
\end{acknowledgments}

%

\vspace{5mm}
\facilities{GBT}



\clearpage
\appendix

\DIFdelbegin 
\DIFdelend 
\DIFaddbegin \begin{longtable*}{cccccccccc}
\DIFaddend \caption{Measured properties of 128 bursts. Column (1): burst index;  Col.(2): Modified Julian dates referenced to infinite frequency at the Solar System barycentre; Col.(3): peak flux density; Col.(4): width; Col.(5): fluence; Col.(6): isotropic equivalent burst energy; Col.(7): dispersion measure; Col.(8): RM obtained by Stokes QU-fitting; Col.(9): degree of linear polarization; Col.(10): degree of circular polarization. }\label{tab:burst}\\\toprule
\endfirsthead
\caption{Continued:}\\\toprule
\endhead
\endfoot
\bottomrule
\endlastfoot
ID & MJD & $S_{\rm peak}$ & $W_{\rm eq}$ & $F$ & $E$ & DM & RM & \% Linear & \% Circular \\ 
      &      &   (Jy)  & (ms) &   (Jy ms)  & ($10^{37}$ erg)  & (pc~cm$^{-3}$) & (rad~m$^{-2}$)\\ 
\midrule
\midrule
1 & 59876.05243363 & $1.32_{-0.03}^{+0.03}$ & $2.40_{-0.05}^{+0.05}$ & $3.2_{-0.1}^{+0.1}$ & $3.1_{-0.1}^{+0.1}$ & $219.0_{-0.7}^{+0.7}$ & $19.9_{-9.8}^{+8.6}$ & $94.9_{-4.7}^{+4.7}$ & $-14.9_{-3.5}^{+3.5}$\\
2 & 59876.05243382 & $1.161_{-0.008}^{+0.008}$ & $10.77_{-0.01}^{+0.01}$ & $12.51_{-0.09}^{+0.09}$ & $39.1_{-0.3}^{+0.3}$ & $220.6_{-0.3}^{+0.3}$ & $3.3_{-1.3}^{+1.3}$ & $96.03_{-0.94}^{+0.94}$ & $-20.81_{-0.69}^{+0.69}$\\
3 & 59876.05320168 & $0.634_{-0.007}^{+0.007}$ & $6.01_{-0.02}^{+0.02}$ & $3.81_{-0.05}^{+0.05}$ & $11.5_{-0.2}^{+0.2}$ & $221_{-3}^{+3}$ & $0.1_{-2.9}^{+2.8}$ & $101.0_{-2.1}^{+2.1}$ & $24.8_{-1.5}^{+1.5}$\\
4 & 59876.05499184 & $1.34_{-0.02}^{+0.02}$ & $5.42_{-0.03}^{+0.03}$ & $7.3_{-0.2}^{+0.2}$ & $7.2_{-0.2}^{+0.2}$ & $220_{-2}^{+2}$ & $-4.5_{-7.6}^{+7.5}$ & $97.4_{-4.5}^{+4.5}$ & $-0.2_{-3.2}^{+3.2}$\\
5 & 59876.05502257 & $1.15_{-0.02}^{+0.02}$ & $4.76_{-0.02}^{+0.02}$ & $5.46_{-0.06}^{+0.06}$ & $9.7_{-0.1}^{+0.1}$ & $221_{-5}^{+5}$ & $0.3_{-3.2}^{+3.2}$ & $98.7_{-1.9}^{+1.9}$ & $-8.7_{-1.3}^{+1.3}$\\
6 & 59876.05637282 & $0.55_{-0.01}^{+0.01}$ & $2.17_{-0.04}^{+0.04}$ & $1.19_{-0.04}^{+0.04}$ & $2.13_{-0.07}^{+0.07}$ & $220.7_{-0.4}^{+0.4}$ & $23.5_{-9.8}^{+9.7}$ & $103.1_{-5.6}^{+5.6}$ & $11.5_{-3.9}^{+3.9}$\\
7 & 59876.05704149 & $0.367_{-0.007}^{+0.007}$ & $4.81_{-0.04}^{+0.04}$ & $1.77_{-0.06}^{+0.06}$ & $4.8_{-0.2}^{+0.2}$ & $220_{-4}^{+4}$ & $9.5_{-5.7}^{+5.6}$ & $96.1_{-4.3}^{+4.3}$ & $16.3_{-3.1}^{+3.1}$\\
8 & 59876.05733743 & $0.20_{-0.01}^{+0.01}$ & $3.3_{-0.1}^{+0.1}$ & $0.67_{-0.06}^{+0.06}$ & $1.7_{-0.1}^{+0.1}$ & $222.5_{-0.6}^{+0.6}$ & $-16_{-34}^{+38}$ & $71_{-12}^{+12}$ & $28_{-10}^{+10}$\\
9 & 59876.05739151 & $0.34_{-0.03}^{+0.03}$ & $5.4_{-0.1}^{+0.1}$ & $1.9_{-0.2}^{+0.2}$ & $1.8_{-0.2}^{+0.2}$ & $219_{-3}^{+3}$ & $-24_{-19}^{+24}$ & $80_{-12}^{+12}$ & $18.7_{-9.6}^{+9.6}$\\
10 & 59876.05748434 & $0.70_{-0.01}^{+0.01}$ & $5.44_{-0.04}^{+0.04}$ & $3.8_{-0.1}^{+0.1}$ & $4.9_{-0.2}^{+0.2}$ & $219_{-2}^{+2}$ & $15.5_{-9.2}^{+9.1}$ & $94.9_{-4.3}^{+4.3}$ & $-12.4_{-3.1}^{+3.1}$\\
11 & 59876.05848084 & $0.23_{-0.01}^{+0.01}$ & $5.69_{-0.08}^{+0.08}$ & $1.31_{-0.09}^{+0.09}$ & $2.2_{-0.2}^{+0.2}$ & $218_{-2}^{+2}$ & $1_{-17}^{+16}$ & $68.4_{-7.0}^{+7.0}$ & $-17.6_{-5.9}^{+5.9}$\\
12 & 59876.05864244 & $0.138_{-0.005}^{+0.005}$ & $7.74_{-0.07}^{+0.07}$ & $1.07_{-0.07}^{+0.07}$ & $6.4_{-0.4}^{+0.4}$ & $221_{-5}^{+5}$ & $-7.5_{-6.3}^{+6.4}$ & $78.5_{-5.2}^{+5.2}$ & $-30.1_{-4.2}^{+4.2}$\\
13 & 59876.05926691 & $0.23_{-0.03}^{+0.03}$ & $6.0_{-0.2}^{+0.2}$ & $1.4_{-0.2}^{+0.2}$ & $0.9_{-0.1}^{+0.1}$ & $219_{-1}^{+1}$ & $121_{-44}^{+34}$ & $91_{-19}^{+19}$ & $52_{-16}^{+16}$\\
14 & 59876.05938959 & $0.34_{-0.02}^{+0.02}$ & $3.16_{-0.07}^{+0.07}$ & $1.08_{-0.05}^{+0.05}$ & $1.83_{-0.09}^{+0.09}$ & $218_{-2}^{+2}$ & $45_{-30}^{+29}$ & $92.8_{-7.7}^{+7.7}$ & $18.5_{-5.7}^{+5.7}$\\
15 & 59876.05966091 & $0.29_{-0.01}^{+0.01}$ & $7.07_{-0.07}^{+0.07}$ & $2.0_{-0.1}^{+0.1}$ & $2.5_{-0.1}^{+0.1}$ & $220_{-6}^{+6}$ & $5_{-14}^{+14}$ & $103.3_{-6.8}^{+6.8}$ & $4.9_{-4.7}^{+4.7}$\\
16 & 59876.06011839 & $0.38_{-0.01}^{+0.01}$ & $6.47_{-0.06}^{+0.06}$ & $2.5_{-0.1}^{+0.1}$ & $4.0_{-0.2}^{+0.2}$ & $221_{-2}^{+2}$ & $3_{-10}^{+10}$ & $90.4_{-5.5}^{+5.5}$ & $2.9_{-4.1}^{+4.1}$\\
17 & 59876.06012005 & $0.51_{-0.02}^{+0.02}$ & $7.82_{-0.07}^{+0.07}$ & $4.0_{-0.2}^{+0.2}$ & $2.7_{-0.2}^{+0.2}$ & $219_{-2}^{+2}$ & $10_{-17}^{+17}$ & $93.3_{-8.5}^{+8.5}$ & $28.3_{-6.5}^{+6.5}$\\
18 & 59876.06059119 & $0.42_{-0.04}^{+0.04}$ & $4.2_{-0.1}^{+0.1}$ & $1.8_{-0.2}^{+0.2}$ & $1.7_{-0.2}^{+0.2}$ & $219.0_{-0.7}^{+0.7}$ & $-9_{-23}^{+24}$ & $77_{-15}^{+15}$ & $-30_{-12}^{+12}$\\
19 & 59876.06252151 & $0.38_{-0.01}^{+0.01}$ & $5.86_{-0.04}^{+0.04}$ & $2.26_{-0.06}^{+0.06}$ & $7.4_{-0.2}^{+0.2}$ & $218.8_{-0.2}^{+0.2}$ & $-5.3_{-3.6}^{+3.7}$ & $96.0_{-3.3}^{+3.3}$ & $13.5_{-2.4}^{+2.4}$\\
20 & 59876.06275827 & $0.16_{-0.01}^{+0.01}$ & $4.6_{-0.1}^{+0.1}$ & $0.73_{-0.06}^{+0.06}$ & $2.4_{-0.2}^{+0.2}$ & $219_{-6}^{+6}$ & $20_{-20}^{+18}$ & $62.5_{-9.9}^{+9.9}$ & $28.0_{-8.7}^{+8.7}$\\
21 & 59876.06391244 & $0.36_{-0.01}^{+0.01}$ & $5.91_{-0.05}^{+0.05}$ & $2.14_{-0.08}^{+0.08}$ & $3.8_{-0.1}^{+0.1}$ & $221_{-6}^{+6}$ & $14_{-11}^{+10}$ & $82.1_{-5.4}^{+5.4}$ & $6.9_{-4.2}^{+4.2}$\\
22 & 59876.06478398 & $0.17_{-0.02}^{+0.02}$ & $5.3_{-0.1}^{+0.1}$ & $0.89_{-0.09}^{+0.09}$ & $1.5_{-0.2}^{+0.2}$ & $221_{-5}^{+5}$ & $25_{-25}^{+23}$ & $81_{-13}^{+13}$ & $10_{-10}^{+10}$\\
23 & 59876.06502637 & $0.35_{-0.01}^{+0.01}$ & $3.90_{-0.05}^{+0.05}$ & $1.35_{-0.04}^{+0.04}$ & $4.1_{-0.1}^{+0.1}$ & $221.8_{-0.9}^{+0.9}$ & $-20.9_{-8.9}^{+8.8}$ & $85.7_{-6.0}^{+6.0}$ & $23.7_{-4.7}^{+4.7}$\\
24 & 59876.06502715 & $0.28_{-0.01}^{+0.01}$ & $8.12_{-0.07}^{+0.07}$ & $2.2_{-0.1}^{+0.1}$ & $9.5_{-0.5}^{+0.5}$ & $220_{-5}^{+5}$ & $13.9_{-3.5}^{+3.5}$ & $92.8_{-5.0}^{+5.0}$ & $8.5_{-3.7}^{+3.7}$\\
25 & 59876.06533531 & $0.110_{-0.004}^{+0.004}$ & $10.74_{-0.08}^{+0.08}$ & $1.18_{-0.09}^{+0.09}$ & $3.8_{-0.3}^{+0.3}$ & $219.7_{-0.2}^{+0.2}$ & $-72_{-15}^{+19}$ & $69.1_{-8.3}^{+8.3}$ & $9.3_{-6.9}^{+6.9}$\\
26 & 59876.06533873 & $0.42_{-0.03}^{+0.03}$ & $6.5_{-0.1}^{+0.1}$ & $2.7_{-0.2}^{+0.2}$ & $2.7_{-0.2}^{+0.2}$ & $219_{-5}^{+5}$ & $45_{-14}^{+12}$ & $80.2_{-9.0}^{+9.0}$ & $12.3_{-7.1}^{+7.1}$\\
27 & 59876.06623892 & $0.59_{-0.01}^{+0.01}$ & $3.63_{-0.03}^{+0.03}$ & $2.15_{-0.05}^{+0.05}$ & $6.6_{-0.2}^{+0.2}$ & $222_{-3}^{+3}$ & $2.9_{-4.2}^{+4.2}$ & $89.4_{-3.1}^{+3.1}$ & $22.3_{-2.4}^{+2.4}$\\
28 & 59876.06630152 & $0.13_{-0.02}^{+0.02}$ & $3.1_{-0.2}^{+0.2}$ & $0.42_{-0.08}^{+0.08}$ & $0.6_{-0.1}^{+0.1}$ & $222.6_{-0.9}^{+0.9}$ & $-98_{-35}^{+45}$ & $97_{-22}^{+22}$ & $-3_{-16}^{+16}$\\
29 & 59876.06635495 & $0.621_{-0.009}^{+0.009}$ & $4.93_{-0.02}^{+0.02}$ & $3.06_{-0.05}^{+0.05}$ & $9.6_{-0.1}^{+0.1}$ & $222.1_{-0.9}^{+0.9}$ & $-2.7_{-2.8}^{+2.9}$ & $98.5_{-2.3}^{+2.3}$ & $22.6_{-1.6}^{+1.6}$\\
30 & 59876.06635518 & $0.646_{-0.008}^{+0.008}$ & $3.26_{-0.02}^{+0.02}$ & $2.10_{-0.04}^{+0.04}$ & $6.6_{-0.1}^{+0.1}$ & $220.5_{-0.4}^{+0.4}$ & $-1.3_{-3.5}^{+3.6}$ & $92.0_{-2.6}^{+2.6}$ & $21.1_{-2.0}^{+2.0}$\\
31 & 59876.06660592 & $0.333_{-0.004}^{+0.004}$ & $5.65_{-0.02}^{+0.02}$ & $1.88_{-0.04}^{+0.04}$ & $12.1_{-0.2}^{+0.2}$ & $222_{-3}^{+3}$ & $-5.1_{-3.1}^{+3.5}$ & $95.3_{-3.1}^{+3.1}$ & $5.9_{-2.3}^{+2.3}$\\
32 & 59876.06660619 & $0.202_{-0.003}^{+0.003}$ & $12.96_{-0.03}^{+0.03}$ & $2.62_{-0.07}^{+0.07}$ & $15.6_{-0.4}^{+0.4}$ & $221.7_{-0.6}^{+0.6}$ & $-6.9_{-4.0}^{+4.3}$ & $94.5_{-3.9}^{+3.9}$ & $-3.2_{-2.8}^{+2.8}$\\
33 & 59876.06701317 & $0.218_{-0.007}^{+0.007}$ & $5.23_{-0.05}^{+0.05}$ & $1.14_{-0.05}^{+0.05}$ & $3.7_{-0.2}^{+0.2}$ & $216.6_{-0.7}^{+0.7}$ & $-8_{-13}^{+13}$ & $94.2_{-5.8}^{+5.8}$ & $20.2_{-4.3}^{+4.3}$\\
34 & 59876.06793962 & $0.469_{-0.006}^{+0.006}$ & $5.05_{-0.02}^{+0.02}$ & $2.37_{-0.05}^{+0.05}$ & $12.1_{-0.2}^{+0.2}$ & $220_{-4}^{+4}$ & $5.1_{-3.4}^{+3.3}$ & $95.3_{-2.5}^{+2.5}$ & $24.3_{-1.8}^{+1.8}$\\
35 & 59876.06895464 & $0.25_{-0.01}^{+0.01}$ & $5.58_{-0.05}^{+0.05}$ & $1.41_{-0.05}^{+0.05}$ & $4.7_{-0.2}^{+0.2}$ & $219_{-4}^{+4}$ & $-6_{-13}^{+16}$ & $76.4_{-4.9}^{+4.9}$ & $42.3_{-4.3}^{+4.3}$\\
36 & 59876.06899501 & $0.15_{-0.01}^{+0.01}$ & $6.3_{-0.1}^{+0.1}$ & $0.95_{-0.08}^{+0.08}$ & $2.6_{-0.2}^{+0.2}$ & $220_{-2}^{+2}$ & $-14_{-25}^{+27}$ & $96_{-10}^{+10}$ & $8.5_{-7.5}^{+7.5}$\\
37 & 59876.06996962 & $0.47_{-0.03}^{+0.03}$ & $5.0_{-0.1}^{+0.1}$ & $2.3_{-0.2}^{+0.2}$ & $2.3_{-0.2}^{+0.2}$ & $218_{-5}^{+5}$ & $-8_{-19}^{+18}$ & $95_{-11}^{+11}$ & $28.5_{-8.0}^{+8.0}$\\
38 & 59876.06996992 & $0.42_{-0.02}^{+0.02}$ & $6.66_{-0.09}^{+0.09}$ & $2.8_{-0.2}^{+0.2}$ & $2.8_{-0.2}^{+0.2}$ & $219_{-6}^{+6}$ & $-59_{-17}^{+23}$ & $82.6_{-9.1}^{+9.1}$ & $-26.6_{-7.2}^{+7.2}$\\
39 & 59876.07042201 & $0.398_{-0.004}^{+0.004}$ & $10.38_{-0.02}^{+0.02}$ & $4.14_{-0.08}^{+0.08}$ & $13.3_{-0.3}^{+0.3}$ & $219.5_{-0.5}^{+0.5}$ & $0.3_{-3.4}^{+3.6}$ & $87.5_{-2.5}^{+2.5}$ & $0.4_{-1.9}^{+1.9}$\\
40 & 59876.07133003 & $0.154_{-0.005}^{+0.005}$ & $10.81_{-0.05}^{+0.05}$ & $1.67_{-0.05}^{+0.05}$ & $13.0_{-0.4}^{+0.4}$ & $220_{-5}^{+5}$ & $15.1_{-4.9}^{+3.7}$ & $51.6_{-3.3}^{+3.3}$ & $32.9_{-3.1}^{+3.1}$\\
41 & 59876.07151962 & $0.206_{-0.008}^{+0.008}$ & $2.6_{-0.1}^{+0.1}$ & $0.53_{-0.05}^{+0.05}$ & $0.78_{-0.08}^{+0.08}$ & $222_{-1}^{+1}$ & $81_{-154}^{+74}$ & $100_{-16}^{+16}$ & $29_{-12}^{+12}$\\
42 & 59876.07174466 & $0.215_{-0.007}^{+0.007}$ & $6.36_{-0.05}^{+0.05}$ & $1.37_{-0.05}^{+0.05}$ & $8.5_{-0.3}^{+0.3}$ & $222_{-4}^{+4}$ & $6.1_{-5.9}^{+5.7}$ & $92.0_{-4.6}^{+4.6}$ & $15.1_{-3.5}^{+3.5}$\\
43 & 59876.07216021 & $0.54_{-0.04}^{+0.04}$ & $2.5_{-0.1}^{+0.1}$ & $1.4_{-0.1}^{+0.1}$ & $1.3_{-0.1}^{+0.1}$ & $219_{-5}^{+5}$ & $-32_{-21}^{+24}$ & $97_{-12}^{+12}$ & $22.7_{-8.8}^{+8.8}$\\
44 & 59876.07261354 & $0.105_{-0.007}^{+0.007}$ & $7.1_{-0.1}^{+0.1}$ & $0.75_{-0.06}^{+0.06}$ & $3.6_{-0.3}^{+0.3}$ & $221_{-1}^{+1}$ & $27_{-13}^{+9}$ & $84.1_{-8.8}^{+8.8}$ & $8.7_{-6.8}^{+6.8}$\\
45 & 59876.07491485 & $0.27_{-0.01}^{+0.01}$ & $7.08_{-0.07}^{+0.07}$ & $1.90_{-0.09}^{+0.09}$ & $8.5_{-0.4}^{+0.4}$ & $221_{-4}^{+4}$ & $6.7_{-4.0}^{+3.6}$ & $73.6_{-5.7}^{+5.7}$ & $52.7_{-5.2}^{+5.2}$\\
46 & 59876.07499691 & $0.44_{-0.01}^{+0.01}$ & $9.41_{-0.07}^{+0.07}$ & $4.1_{-0.3}^{+0.3}$ & $4.1_{-0.3}^{+0.3}$ & $220_{-2}^{+2}$ & $12_{-13}^{+14}$ & $71.7_{-6.7}^{+6.7}$ & $-57.8_{-6.3}^{+6.3}$\\
47 & 59876.07542136 & $0.38_{-0.01}^{+0.01}$ & $5.03_{-0.06}^{+0.06}$ & $1.90_{-0.09}^{+0.09}$ & $4.2_{-0.2}^{+0.2}$ & $218_{-4}^{+4}$ & $-14_{-11}^{+11}$ & $101.0_{-5.9}^{+5.9}$ & $-9.5_{-4.2}^{+4.2}$\\
48 & 59876.07601342 & $0.20_{-0.01}^{+0.01}$ & $4.70_{-0.09}^{+0.09}$ & $0.95_{-0.06}^{+0.06}$ & $3.1_{-0.2}^{+0.2}$ & $220_{-4}^{+4}$ & $16_{-11}^{+10}$ & $68.8_{-6.7}^{+6.7}$ & $15.5_{-5.6}^{+5.6}$\\
49 & 59876.07635757 & $0.42_{-0.01}^{+0.01}$ & $3.59_{-0.05}^{+0.05}$ & $1.51_{-0.07}^{+0.07}$ & $6.4_{-0.3}^{+0.3}$ & $220_{-4}^{+4}$ & $1_{-12}^{+10}$ & $76.5_{-5.2}^{+5.2}$ & $34.8_{-4.4}^{+4.4}$\\
50 & 59876.07662495 & $0.393_{-0.005}^{+0.005}$ & $10.56_{-0.02}^{+0.02}$ & $4.15_{-0.06}^{+0.06}$ & $24.9_{-0.4}^{+0.4}$ & $224_{-1}^{+1}$ & $2.5_{-2.0}^{+2.0}$ & $86.8_{-2.0}^{+2.0}$ & $22.2_{-1.5}^{+1.5}$\\
51 & 59876.07788136 & $0.19_{-0.01}^{+0.01}$ & $4.57_{-0.09}^{+0.09}$ & $0.89_{-0.05}^{+0.05}$ & $2.4_{-0.1}^{+0.1}$ & $219_{-5}^{+5}$ & $16_{-14}^{+13}$ & $107.1_{-9.9}^{+9.9}$ & $16.7_{-6.9}^{+6.9}$\\
52 & 59876.07967444 & $3.274_{-0.009}^{+0.009}$ & $2.353_{-0.004}^{+0.004}$ & $7.70_{-0.03}^{+0.03}$ & $59.7_{-0.2}^{+0.2}$ & $221.68_{-0.03}^{+0.03}$ & $2.10_{-0.50}^{+0.50}$ & $94.41_{-0.46}^{+0.46}$ & $29.88_{-0.35}^{+0.35}$\\
53 & 59876.07968784 & $0.39_{-0.04}^{+0.04}$ & $3.8_{-0.1}^{+0.1}$ & $1.5_{-0.2}^{+0.2}$ & $1.4_{-0.1}^{+0.1}$ & $221_{-4}^{+4}$ & $33_{-41}^{+29}$ & $77_{-13}^{+13}$ & $-22_{-10}^{+10}$\\
54 & 59876.08128268 & $0.29_{-0.02}^{+0.02}$ & $7.99_{-0.09}^{+0.09}$ & $2.3_{-0.1}^{+0.1}$ & $7.7_{-0.4}^{+0.4}$ & $219_{-2}^{+2}$ & $-8.1_{-4.6}^{+4.2}$ & $73.4_{-6.2}^{+6.2}$ & $8.5_{-5.0}^{+5.0}$\\
55 & 59876.08167999 & $0.21_{-0.01}^{+0.01}$ & $5.34_{-0.09}^{+0.09}$ & $1.12_{-0.08}^{+0.08}$ & $1.7_{-0.1}^{+0.1}$ & $218_{-2}^{+2}$ & $-54_{-70}^{+81}$ & $73.4_{-8.4}^{+8.4}$ & $10.8_{-6.8}^{+6.8}$\\
56 & 59876.08179005 & $0.15_{-0.02}^{+0.02}$ & $8.5_{-0.2}^{+0.2}$ & $1.3_{-0.2}^{+0.2}$ & $4.5_{-0.6}^{+0.6}$ & $219_{-6}^{+6}$ & $1_{-8}^{+11}$ & $74_{-11}^{+11}$ & $32.7_{-9.4}^{+9.4}$\\
57 & 59876.08193785 & $0.550_{-0.008}^{+0.008}$ & $6.54_{-0.02}^{+0.02}$ & $3.60_{-0.07}^{+0.07}$ & $19.8_{-0.4}^{+0.4}$ & $221.3_{-0.4}^{+0.4}$ & $-3.5_{-3.1}^{+2.9}$ & $77.3_{-2.1}^{+2.1}$ & $27.6_{-1.7}^{+1.7}$\\
58 & 59876.0821865 & $0.55_{-0.03}^{+0.03}$ & $4.54_{-0.09}^{+0.09}$ & $2.5_{-0.2}^{+0.2}$ & $2.5_{-0.2}^{+0.2}$ & $221_{-3}^{+3}$ & $-22_{-14}^{+14}$ & $86.9_{-8.0}^{+8.0}$ & $26.1_{-6.3}^{+6.3}$\\
59 & 59876.08337069 & $0.31_{-0.01}^{+0.01}$ & $5.85_{-0.06}^{+0.06}$ & $1.84_{-0.08}^{+0.08}$ & $9.3_{-0.4}^{+0.4}$ & $221.0_{-0.6}^{+0.6}$ & $8.1_{-8.8}^{+5.6}$ & $89.8_{-5.7}^{+5.7}$ & $8.3_{-4.3}^{+4.3}$\\
60 & 59876.08347388 & $0.088_{-0.009}^{+0.009}$ & $4.9_{-0.2}^{+0.2}$ & $0.43_{-0.05}^{+0.05}$ & $1.4_{-0.2}^{+0.2}$ & $223.3_{-0.8}^{+0.8}$ & $-24_{-40}^{+45}$ & $96_{-14}^{+14}$ & $36_{-11}^{+11}$\\
61 & 59876.08401727 & $0.48_{-0.01}^{+0.01}$ & $12.98_{-0.02}^{+0.02}$ & $6.20_{-0.09}^{+0.09}$ & $22.7_{-0.3}^{+0.3}$ & $228_{-3}^{+3}$ & $-15.3_{-4.8}^{+5.1}$ & $74.1_{-2.7}^{+2.7}$ & $18.3_{-2.2}^{+2.2}$\\
62 & 59876.08461302 & $0.128_{-0.008}^{+0.008}$ & $3.5_{-0.1}^{+0.1}$ & $0.44_{-0.05}^{+0.05}$ & $2.0_{-0.2}^{+0.2}$ & $219_{-3}^{+3}$ & $72_{-15}^{+16}$ & $75.8_{-9.8}^{+9.8}$ & $13.7_{-7.9}^{+7.9}$\\
63 & 59876.0852808 & $0.47_{-0.03}^{+0.03}$ & $6.6_{-0.1}^{+0.1}$ & $3.1_{-0.3}^{+0.3}$ & $3.1_{-0.3}^{+0.3}$ & $219_{-2}^{+2}$ & $5_{-20}^{+17}$ & $60.0_{-8.0}^{+8.0}$ & $27.6_{-7.1}^{+7.1}$\\
64 & 59876.08545989 & $0.148_{-0.007}^{+0.007}$ & $8.80_{-0.07}^{+0.07}$ & $1.30_{-0.06}^{+0.06}$ & $8.0_{-0.4}^{+0.4}$ & $220_{-5}^{+5}$ & $0.1_{-6.2}^{+5.6}$ & $76.2_{-5.5}^{+5.5}$ & $23.8_{-4.5}^{+4.5}$\\
65 & 59876.08546189 & $0.67_{-0.04}^{+0.04}$ & $4.43_{-0.08}^{+0.08}$ & $3.0_{-0.1}^{+0.1}$ & $3.0_{-0.1}^{+0.1}$ & $218_{-4}^{+4}$ & $-4_{-11}^{+11}$ & $92.0_{-6.3}^{+6.3}$ & $17.3_{-4.7}^{+4.7}$\\
66 & 59876.08627925 & $0.221_{-0.005}^{+0.005}$ & $6.04_{-0.04}^{+0.04}$ & $1.33_{-0.04}^{+0.04}$ & $5.8_{-0.2}^{+0.2}$ & $219_{-5}^{+5}$ & $4.3_{-8.8}^{+7.8}$ & $98.3_{-6.2}^{+6.2}$ & $33.5_{-4.7}^{+4.7}$\\
67 & 59876.08650527 & $0.38_{-0.01}^{+0.01}$ & $5.25_{-0.03}^{+0.03}$ & $1.97_{-0.04}^{+0.04}$ & $6.5_{-0.1}^{+0.1}$ & $218_{-4}^{+4}$ & $-14.7_{-5.5}^{+6.4}$ & $75.2_{-4.1}^{+4.1}$ & $53.1_{-3.7}^{+3.7}$\\
68 & 59876.08669388 & $0.30_{-0.02}^{+0.02}$ & $3.63_{-0.09}^{+0.09}$ & $1.08_{-0.07}^{+0.07}$ & $1.08_{-0.07}^{+0.07}$ & $220_{-2}^{+2}$ & $-182_{-73}^{+72}$ & $69.2_{-9.1}^{+9.1}$ & $21.7_{-7.6}^{+7.6}$\\
69 & 59876.0871982 & $0.217_{-0.008}^{+0.008}$ & $5.13_{-0.06}^{+0.06}$ & $1.11_{-0.05}^{+0.05}$ & $3.1_{-0.1}^{+0.1}$ & $219_{-5}^{+5}$ & $8.3_{-7.6}^{+7.2}$ & $90.8_{-5.9}^{+5.9}$ & $25.5_{-4.5}^{+4.5}$\\
70 & 59876.08759098 & $0.143_{-0.004}^{+0.004}$ & $15.37_{-0.04}^{+0.04}$ & $2.20_{-0.07}^{+0.07}$ & $14.8_{-0.5}^{+0.5}$ & $222_{-2}^{+2}$ & $-11.1_{-4.8}^{+5.0}$ & $86.5_{-4.0}^{+4.0}$ & $27.5_{-3.1}^{+3.1}$\\
71 & 59876.08770016 & $0.215_{-0.005}^{+0.005}$ & $5.18_{-0.04}^{+0.04}$ & $1.12_{-0.04}^{+0.04}$ & $5.0_{-0.2}^{+0.2}$ & $222_{-1}^{+1}$ & $-16_{-10}^{+11}$ & $78.9_{-5.1}^{+5.1}$ & $22.0_{-4.1}^{+4.1}$\\
72 & 59876.08801454 & $0.558_{-0.004}^{+0.004}$ & $8.73_{-0.01}^{+0.01}$ & $4.88_{-0.06}^{+0.06}$ & $27.9_{-0.3}^{+0.3}$ & $219.6_{-0.3}^{+0.3}$ & $8.5_{-1.5}^{+1.4}$ & $94.5_{-1.7}^{+1.7}$ & $33.1_{-1.3}^{+1.3}$\\
73 & 59876.0884081 & $0.13_{-0.01}^{+0.01}$ & $3.4_{-0.1}^{+0.1}$ & $0.44_{-0.05}^{+0.05}$ & $1.1_{-0.1}^{+0.1}$ & $221.0_{-0.4}^{+0.4}$ & $-40_{-26}^{+24}$ & $91_{-16}^{+16}$ & $26_{-12}^{+12}$\\
74 & 59876.08840859 & $0.13_{-0.02}^{+0.02}$ & $3.7_{-0.2}^{+0.2}$ & $0.48_{-0.08}^{+0.08}$ & $0.7_{-0.1}^{+0.1}$ & $223.0_{-0.5}^{+0.5}$ & $-36_{-32}^{+34}$ & $71_{-16}^{+16}$ & $32_{-14}^{+14}$\\
75 & 59876.08857414 & $0.42_{-0.01}^{+0.01}$ & $7.56_{-0.03}^{+0.03}$ & $3.20_{-0.07}^{+0.07}$ & $10.6_{-0.2}^{+0.2}$ & $219.7_{-0.6}^{+0.6}$ & $12.9_{-4.0}^{+3.7}$ & $82.2_{-2.9}^{+2.9}$ & $21.8_{-2.3}^{+2.3}$\\
76 & 59876.08860183 & $1.82_{-0.01}^{+0.01}$ & $2.92_{-0.01}^{+0.01}$ & $5.33_{-0.04}^{+0.04}$ & $14.3_{-0.1}^{+0.1}$ & $219_{-4}^{+4}$ & $3.3_{-2.1}^{+2.0}$ & $88.3_{-1.2}^{+1.2}$ & $29.88_{-0.96}^{+0.96}$\\
77 & 59876.08873233 & $0.106_{-0.006}^{+0.006}$ & $3.2_{-0.2}^{+0.2}$ & $0.34_{-0.05}^{+0.05}$ & $1.3_{-0.2}^{+0.2}$ & $216.5_{-0.2}^{+0.2}$ & $67_{-46}^{+30}$ & $123_{-22}^{+22}$ & $22_{-14}^{+14}$\\
78 & 59876.08944327 & $0.10_{-0.01}^{+0.01}$ & $3.8_{-0.2}^{+0.2}$ & $0.37_{-0.05}^{+0.05}$ & $1.0_{-0.1}^{+0.1}$ & $217.7_{-0.3}^{+0.3}$ & $16_{-29}^{+35}$ & $66_{-11}^{+11}$ & $8.6_{-9.0}^{+9.0}$\\
79 & 59876.08944348 & $0.283_{-0.008}^{+0.008}$ & $7.22_{-0.04}^{+0.04}$ & $2.04_{-0.05}^{+0.05}$ & $10.4_{-0.3}^{+0.3}$ & $223.4_{-0.6}^{+0.6}$ & $-5.3_{-3.0}^{+3.2}$ & $74.1_{-3.0}^{+3.0}$ & $30.9_{-2.6}^{+2.6}$\\
80 & 59876.0895257 & $0.34_{-0.01}^{+0.01}$ & $6.87_{-0.05}^{+0.05}$ & $2.30_{-0.09}^{+0.09}$ & $8.9_{-0.3}^{+0.3}$ & $221_{-2}^{+2}$ & $9.3_{-4.2}^{+3.8}$ & $91.5_{-5.6}^{+5.6}$ & $-41.1_{-4.4}^{+4.4}$\\
81 & 59876.09013786 & $0.43_{-0.01}^{+0.01}$ & $7.35_{-0.04}^{+0.04}$ & $3.16_{-0.07}^{+0.07}$ & $16.2_{-0.4}^{+0.4}$ & $218_{-1}^{+1}$ & $-5.1_{-3.9}^{+3.9}$ & $99.3_{-3.0}^{+3.0}$ & $-1.0_{-2.2}^{+2.2}$\\
82 & 59876.09114072 & $2.809_{-0.008}^{+0.008}$ & $2.723_{-0.005}^{+0.005}$ & $7.65_{-0.03}^{+0.03}$ & $37.2_{-0.1}^{+0.1}$ & $219.9_{-0.4}^{+0.4}$ & $2.1_{-0.9}^{+0.9}$ & $92.98_{-0.55}^{+0.55}$ & $10.80_{-0.41}^{+0.41}$\\
83 & 59876.09135036 & $2.342_{-0.006}^{+0.006}$ & $2.317_{-0.006}^{+0.006}$ & $5.43_{-0.03}^{+0.03}$ & $28.5_{-0.2}^{+0.2}$ & $220.9_{-0.1}^{+0.1}$ & $0.5_{-1.0}^{+1.0}$ & $97.01_{-0.87}^{+0.87}$ & $21.04_{-0.64}^{+0.64}$\\
84 & 59876.09139437 & $0.57_{-0.01}^{+0.01}$ & $4.31_{-0.03}^{+0.03}$ & $2.45_{-0.05}^{+0.05}$ & $9.5_{-0.2}^{+0.2}$ & $224.1_{-0.9}^{+0.9}$ & $-6.3_{-3.2}^{+3.5}$ & $84.7_{-2.7}^{+2.7}$ & $51.4_{-2.3}^{+2.3}$\\
85 & 59876.09165954 & $0.57_{-0.02}^{+0.02}$ & $4.49_{-0.08}^{+0.08}$ & $2.6_{-0.2}^{+0.2}$ & $2.5_{-0.2}^{+0.2}$ & $219_{-2}^{+2}$ & $-65_{-14}^{+16}$ & $64.4_{-7.7}^{+7.7}$ & $-7.7_{-6.5}^{+6.5}$\\
86 & 59876.09166083 & $0.34_{-0.01}^{+0.01}$ & $7.5_{-0.1}^{+0.1}$ & $2.5_{-0.2}^{+0.2}$ & $2.5_{-0.2}^{+0.2}$ & $221_{-2}^{+2}$ & $31_{-23}^{+21}$ & $89_{-11}^{+11}$ & $35.1_{-9.0}^{+9.0}$\\
87 & 59876.09272137 & $0.200_{-0.004}^{+0.004}$ & $7.04_{-0.04}^{+0.04}$ & $1.41_{-0.04}^{+0.04}$ & $9.4_{-0.3}^{+0.3}$ & $220_{-4}^{+4}$ & $6.7_{-4.8}^{+4.4}$ & $90.4_{-4.2}^{+4.2}$ & $33.9_{-3.3}^{+3.3}$\\
88 & 59876.09293592 & $0.302_{-0.005}^{+0.005}$ & $10.68_{-0.03}^{+0.03}$ & $3.22_{-0.07}^{+0.07}$ & $25.3_{-0.5}^{+0.5}$ & $222.5_{-0.2}^{+0.2}$ & $-14.9_{-2.7}^{+3.1}$ & $92.7_{-2.7}^{+2.7}$ & $-17.1_{-2.0}^{+2.0}$\\
89 & 59876.0932935 & $16.90_{-0.02}^{+0.02}$ & $2.700_{-0.005}^{+0.005}$ & $46.60_{-0.06}^{+0.06}$ & $210.0_{-0.3}^{+0.3}$ & $219.7_{-0.3}^{+0.3}$ & $14.50_{-0.60}^{+0.60}$ & $68.60_{-0.30}^{+0.30}$ & $15.80_{-0.20}^{+0.20}$\\
90 & 59876.09329438 & $3.54_{-0.01}^{+0.01}$ & $7.119_{-0.004}^{+0.004}$ & $25.21_{-0.07}^{+0.07}$ & $131.0_{-0.4}^{+0.4}$ & $219.1_{-0.3}^{+0.3}$ & $4.30_{-0.50}^{+0.50}$ & $95.73_{-0.35}^{+0.35}$ & $17.88_{-0.26}^{+0.26}$\\
91 & 59876.09368221 & $0.42_{-0.01}^{+0.01}$ & $11.04_{-0.03}^{+0.03}$ & $4.65_{-0.09}^{+0.09}$ & $25.1_{-0.5}^{+0.5}$ & $224_{-1}^{+1}$ & $9.9_{-3.3}^{+2.8}$ & $85.1_{-2.4}^{+2.4}$ & $40.2_{-2.0}^{+2.0}$\\
92 & 59876.09374917 & $0.34_{-0.03}^{+0.03}$ & $4.9_{-0.1}^{+0.1}$ & $1.7_{-0.2}^{+0.2}$ & $1.6_{-0.2}^{+0.2}$ & $220_{-4}^{+4}$ & $-48_{-29}^{+22}$ & $92_{-13}^{+13}$ & $-17_{-10}^{+10}$\\
93 & 59876.09386193 & $0.290_{-0.009}^{+0.009}$ & $2.52_{-0.05}^{+0.05}$ & $0.73_{-0.03}^{+0.03}$ & $2.8_{-0.1}^{+0.1}$ & $222.1_{-0.5}^{+0.5}$ & $29_{-11}^{+9}$ & $105.8_{-6.6}^{+6.6}$ & $16.7_{-4.6}^{+4.6}$\\
94 & 59876.09439839 & $0.23_{-0.02}^{+0.02}$ & $6.1_{-0.2}^{+0.2}$ & $1.4_{-0.2}^{+0.2}$ & $1.4_{-0.2}^{+0.2}$ & $220_{-2}^{+2}$ & $31_{-31}^{+35}$ & $75_{-15}^{+15}$ & $22_{-12}^{+12}$\\
95 & 59876.09448982 & $0.170_{-0.006}^{+0.006}$ & $4.5_{-0.1}^{+0.1}$ & $0.77_{-0.08}^{+0.08}$ & $1.9_{-0.2}^{+0.2}$ & $219_{-4}^{+4}$ & $17_{-32}^{+37}$ & $123_{-16}^{+16}$ & $40_{-11}^{+11}$\\
96 & 59876.09506406 & $0.396_{-0.007}^{+0.007}$ & $2.95_{-0.03}^{+0.03}$ & $1.17_{-0.02}^{+0.02}$ & $7.5_{-0.2}^{+0.2}$ & $218.8_{-0.2}^{+0.2}$ & $-0.7_{-2.5}^{+2.6}$ & $93.6_{-3.0}^{+3.0}$ & $28.9_{-2.3}^{+2.3}$\\
97 & 59876.09506419 & $0.244_{-0.006}^{+0.006}$ & $6.38_{-0.03}^{+0.03}$ & $1.55_{-0.04}^{+0.04}$ & $10.2_{-0.2}^{+0.2}$ & $214.6_{-0.5}^{+0.5}$ & $-7.7_{-3.2}^{+4.0}$ & $86.9_{-3.2}^{+3.2}$ & $29.0_{-2.5}^{+2.5}$\\
98 & 59876.09588408 & $0.127_{-0.007}^{+0.007}$ & $4.8_{-0.1}^{+0.1}$ & $0.61_{-0.06}^{+0.06}$ & $1.7_{-0.2}^{+0.2}$ & $217.2_{-0.3}^{+0.3}$ & $-70_{-16}^{+31}$ & $59.1_{-9.9}^{+9.9}$ & $13.0_{-8.6}^{+8.6}$\\
99 & 59876.09727359 & $0.17_{-0.01}^{+0.01}$ & $5.3_{-0.1}^{+0.1}$ & $0.89_{-0.09}^{+0.09}$ & $1.8_{-0.2}^{+0.2}$ & $219_{-3}^{+3}$ & $-12_{-18}^{+18}$ & $73_{-11}^{+11}$ & $-8.5_{-8.6}^{+8.6}$\\
100 & 59876.09732127 & $0.28_{-0.01}^{+0.01}$ & $2.25_{-0.08}^{+0.08}$ & $0.62_{-0.04}^{+0.04}$ & $2.0_{-0.1}^{+0.1}$ & $221.2_{-0.3}^{+0.3}$ & $13_{-16}^{+12}$ & $86.1_{-7.8}^{+7.8}$ & $23.4_{-6.1}^{+6.1}$\\
101 & 59876.09817496 & $1.27_{-0.03}^{+0.03}$ & $2.83_{-0.05}^{+0.05}$ & $3.6_{-0.1}^{+0.1}$ & $3.6_{-0.1}^{+0.1}$ & $220_{-4}^{+4}$ & $2.9_{-8.3}^{+8.7}$ & $87.7_{-4.7}^{+4.7}$ & $10.0_{-3.5}^{+3.5}$\\
102 & 59876.09857921 & $0.67_{-0.04}^{+0.04}$ & $6.51_{-0.07}^{+0.07}$ & $4.4_{-0.2}^{+0.2}$ & $2.9_{-0.1}^{+0.1}$ & $219_{-2}^{+2}$ & $26_{-13}^{+12}$ & $102.3_{-6.3}^{+6.3}$ & $22.6_{-4.5}^{+4.5}$\\
103 & 59876.09926156 & $0.22_{-0.01}^{+0.01}$ & $3.6_{-0.1}^{+0.1}$ & $0.80_{-0.06}^{+0.06}$ & $1.13_{-0.08}^{+0.08}$ & $219_{-6}^{+6}$ & $28_{-34}^{+32}$ & $96_{-13}^{+13}$ & $56_{-11}^{+11}$\\
104 & 59876.0992619 & $0.97_{-0.01}^{+0.01}$ & $1.17_{-0.03}^{+0.03}$ & $1.14_{-0.03}^{+0.03}$ & $3.33_{-0.08}^{+0.08}$ & $219.4_{-0.1}^{+0.1}$ & $-2.1_{-3.7}^{+3.7}$ & $98.5_{-3.2}^{+3.2}$ & $-8.4_{-2.3}^{+2.3}$\\
105 & 59876.09926287 & $0.315_{-0.007}^{+0.007}$ & $6.94_{-0.04}^{+0.04}$ & $2.19_{-0.07}^{+0.07}$ & $4.2_{-0.1}^{+0.1}$ & $225_{-2}^{+2}$ & $-23.1_{-9.0}^{+9.5}$ & $92.2_{-5.4}^{+5.4}$ & $22.0_{-4.1}^{+4.1}$\\
106 & 59876.09965238 & $0.35_{-0.01}^{+0.01}$ & $5.05_{-0.07}^{+0.07}$ & $1.8_{-0.1}^{+0.1}$ & $5.8_{-0.4}^{+0.4}$ & $222_{-1}^{+1}$ & $13.5_{-6.4}^{+4.8}$ & $82.4_{-8.0}^{+8.0}$ & $1.7_{-6.2}^{+6.2}$\\
107 & 59876.09981865 & $0.57_{-0.03}^{+0.03}$ & $4.70_{-0.09}^{+0.09}$ & $2.7_{-0.2}^{+0.2}$ & $1.7_{-0.1}^{+0.1}$ & $219_{-2}^{+2}$ & $32_{-20}^{+19}$ & $100_{-10}^{+10}$ & $-2.3_{-7.1}^{+7.1}$\\
108 & 59876.09988119 & $0.20_{-0.02}^{+0.02}$ & $4.2_{-0.1}^{+0.1}$ & $0.85_{-0.07}^{+0.07}$ & $2.1_{-0.2}^{+0.2}$ & $221.3_{-0.5}^{+0.5}$ & $45_{-19}^{+17}$ & $63.2_{-8.4}^{+8.4}$ & $17.5_{-7.2}^{+7.2}$\\
109 & 59876.10046173 & $0.183_{-0.007}^{+0.007}$ & $5.63_{-0.08}^{+0.08}$ & $1.03_{-0.07}^{+0.07}$ & $3.0_{-0.2}^{+0.2}$ & $220_{-2}^{+2}$ & $11.3_{-9.5}^{+9.8}$ & $69.8_{-7.2}^{+7.2}$ & $8.8_{-6.0}^{+6.0}$\\
110 & 59876.10056288 & $0.349_{-0.006}^{+0.006}$ & $7.75_{-0.03}^{+0.03}$ & $2.70_{-0.06}^{+0.06}$ & $11.0_{-0.2}^{+0.2}$ & $221_{-4}^{+4}$ & $-4.9_{-3.2}^{+3.1}$ & $97.0_{-3.0}^{+3.0}$ & $23.9_{-2.2}^{+2.2}$\\
111 & 59876.10106218 & $0.63_{-0.01}^{+0.01}$ & $4.21_{-0.02}^{+0.02}$ & $2.64_{-0.04}^{+0.04}$ & $13.4_{-0.2}^{+0.2}$ & $221.5_{-0.5}^{+0.5}$ & $-2.3_{-2.1}^{+2.3}$ & $88.7_{-2.2}^{+2.2}$ & $28.6_{-1.7}^{+1.7}$\\
112 & 59876.1011338 & $0.235_{-0.007}^{+0.007}$ & $2.61_{-0.06}^{+0.06}$ & $0.61_{-0.03}^{+0.03}$ & $2.8_{-0.2}^{+0.2}$ & $220.3_{-0.6}^{+0.6}$ & $-1.1_{-8.0}^{+7.4}$ & $87.9_{-6.9}^{+6.9}$ & $15.6_{-5.2}^{+5.2}$\\
113 & 59876.1046181 & $0.13_{-0.01}^{+0.01}$ & $3.9_{-0.2}^{+0.2}$ & $0.49_{-0.06}^{+0.06}$ & $1.4_{-0.2}^{+0.2}$ & $218.1_{-0.4}^{+0.4}$ & $2_{-23}^{+24}$ & $101_{-13}^{+13}$ & $-2.1_{-9.2}^{+9.2}$\\
114 & 59876.10562406 & $0.36_{-0.02}^{+0.02}$ & $8.8_{-0.1}^{+0.1}$ & $3.2_{-0.3}^{+0.3}$ & $1.9_{-0.2}^{+0.2}$ & $219_{-2}^{+2}$ & $-41_{-21}^{+24}$ & $81.6_{-9.1}^{+9.1}$ & $25.0_{-7.3}^{+7.3}$\\
115 & 59876.10598762 & $0.55_{-0.02}^{+0.02}$ & $6.76_{-0.08}^{+0.08}$ & $3.7_{-0.3}^{+0.3}$ & $3.7_{-0.3}^{+0.3}$ & $221_{-2}^{+2}$ & $10_{-13}^{+11}$ & $93.5_{-7.9}^{+7.9}$ & $26.0_{-5.9}^{+5.9}$\\
116 & 59876.10633532 & $0.446_{-0.005}^{+0.005}$ & $9.43_{-0.02}^{+0.02}$ & $4.21_{-0.06}^{+0.06}$ & $21.3_{-0.3}^{+0.3}$ & $216.0_{-0.6}^{+0.6}$ & $2.7_{-1.7}^{+1.8}$ & $94.0_{-1.8}^{+1.8}$ & $22.8_{-1.4}^{+1.4}$\\
117 & 59876.10697445 & $0.277_{-0.008}^{+0.008}$ & $7.20_{-0.05}^{+0.05}$ & $1.99_{-0.08}^{+0.08}$ & $11.1_{-0.4}^{+0.4}$ & $220.8_{-0.5}^{+0.5}$ & $8.9_{-5.0}^{+5.3}$ & $95.9_{-4.8}^{+4.8}$ & $0.9_{-3.5}^{+3.5}$\\
118 & 59876.10727093 & $0.24_{-0.02}^{+0.02}$ & $3.7_{-0.1}^{+0.1}$ & $0.91_{-0.08}^{+0.08}$ & $3.1_{-0.3}^{+0.3}$ & $220_{-3}^{+3}$ & $7.1_{-7.1}^{+7.5}$ & $108_{-12}^{+12}$ & $7.0_{-8.4}^{+8.4}$\\
119 & 59876.10767258 & $0.71_{-0.01}^{+0.01}$ & $9.51_{-0.02}^{+0.02}$ & $6.76_{-0.07}^{+0.07}$ & $19.2_{-0.2}^{+0.2}$ & $221.7_{-0.7}^{+0.7}$ & $0.7_{-1.8}^{+1.8}$ & $95.3_{-1.4}^{+1.4}$ & $-10.1_{-1.0}^{+1.0}$\\
120 & 59876.1078928 & $0.53_{-0.04}^{+0.04}$ & $9.06_{-0.09}^{+0.09}$ & $4.8_{-0.3}^{+0.3}$ & $3.2_{-0.2}^{+0.2}$ & $219_{-5}^{+5}$ & $6_{-13}^{+12}$ & $88.3_{-6.3}^{+6.3}$ & $1.5_{-4.7}^{+4.7}$\\
121 & 59876.10812154 & $0.242_{-0.009}^{+0.009}$ & $7.16_{-0.06}^{+0.06}$ & $1.74_{-0.07}^{+0.07}$ & $8.9_{-0.4}^{+0.4}$ & $221_{-5}^{+5}$ & $-14.3_{-5.3}^{+5.9}$ & $63.3_{-4.2}^{+4.2}$ & $13.9_{-3.6}^{+3.6}$\\
122 & 59876.10854276 & $0.28_{-0.01}^{+0.01}$ & $3.26_{-0.07}^{+0.07}$ & $0.91_{-0.05}^{+0.05}$ & $2.1_{-0.1}^{+0.1}$ & $219.7_{-0.6}^{+0.6}$ & $9.3_{-9.6}^{+9.0}$ & $116.2_{-8.7}^{+8.7}$ & $2.9_{-5.7}^{+5.7}$\\
123 & 59876.10865704 & $0.149_{-0.009}^{+0.009}$ & $8.8_{-0.1}^{+0.1}$ & $1.3_{-0.1}^{+0.1}$ & $2.1_{-0.2}^{+0.2}$ & $219_{-5}^{+5}$ & $-9_{-31}^{+30}$ & $68.3_{-8.6}^{+8.6}$ & $8.8_{-7.1}^{+7.1}$\\
124 & 59876.10920763 & $0.31_{-0.01}^{+0.01}$ & $4.16_{-0.06}^{+0.06}$ & $1.27_{-0.07}^{+0.07}$ & $2.3_{-0.1}^{+0.1}$ & $220_{-2}^{+2}$ & $-13_{-15}^{+16}$ & $84.0_{-6.3}^{+6.3}$ & $-53.2_{-5.4}^{+5.4}$\\
125 & 59876.11092891 & $0.25_{-0.03}^{+0.03}$ & $8.1_{-0.1}^{+0.1}$ & $2.0_{-0.1}^{+0.1}$ & $9.2_{-0.6}^{+0.6}$ & $218.5_{-0.3}^{+0.3}$ & $-1.1_{-4.3}^{+4.2}$ & $98.2_{-6.6}^{+6.6}$ & $16.8_{-4.8}^{+4.8}$\\
126 & 59876.11093197 & $1.897_{-0.007}^{+0.007}$ & $2.23_{-0.03}^{+0.03}$ & $4.2_{-0.1}^{+0.1}$ & $4.2_{-0.1}^{+0.1}$ & $220_{-1}^{+1}$ & $6_{-13}^{+13}$ & $85.8_{-6.0}^{+6.0}$ & $-2.6_{-4.5}^{+4.5}$\\
127 & 59876.11119868 & $0.36_{-0.01}^{+0.01}$ & $7.30_{-0.04}^{+0.04}$ & $2.63_{-0.06}^{+0.06}$ & $14.7_{-0.3}^{+0.3}$ & $221_{-2}^{+2}$ & $1.3_{-4.3}^{+4.1}$ & $85.3_{-2.7}^{+2.7}$ & $11.1_{-2.1}^{+2.1}$\\
128 & 59876.11122224 & $0.110_{-0.006}^{+0.006}$ & $7.19_{-0.08}^{+0.08}$ & $0.79_{-0.05}^{+0.05}$ & $5.0_{-0.3}^{+0.3}$ & $223.9_{-0.7}^{+0.7}$ & $19_{-11}^{+8}$ & $95.5_{-7.7}^{+7.7}$ & $-20.0_{-5.7}^{+5.7}$\\
\end{longtable*}


\bibliography{sample631}{}
\bibliographystyle{aasjournal}



\end{document}